# Machine Learning-Driven Optimization of TPMS Architected Materials Using Simulated Annealing

Akshansh Mishra[1]

[1]School of Industrial and Information Engineering, Politecnico Di Milano, Milan, Italy

Mail id: akshansh.mishra@mail.polimi.it

**Abstract:** The research paper presents a novel approach to optimizing the tensile stress of Triply Periodic Minimal Surface (TPMS) structures through machine learning and Simulated Annealing (SA). The study evaluates the performance of Random Forest, Decision Tree, and XGBoost models in predicting tensile stress, using a dataset generated from finite element analysis of TPMS models. The objective function minimized the negative R-squared value on the validation set to enhance model accuracy. The SA-XGBoost model outperformed the others, achieving an R-squared ($R^2$) value of 0.96. In contrast, the SA-Random Forest model achieved an $R^2$ of 0.89 while the SA-Decision Tree model exhibited greater fluctuations in validation scores. This demonstrates that the SA-XGBoost model is most effective in capturing the complex relationships within the data. The integration of SA helps in optimizing the hyperparameters of these machine learning models, thereby enhancing their predictive capabilities.

1. Introduction

Triply Periodic Minimal Surfaces (TPMS) became known as an important topic of research due to their distinct structure and functional features [1-4]. These surfaces are distinguished by their three-dimensional periodicity and minimal surface area for a given volume, resulting in multiple favorable features that can be used in a variety of technical applications. The relevance of TPMS structures stems from their highly organized pore networks, which provide superior mechanical and thermal properties when compared to traditional materials [5-7]. These structures exist in both natural and synthetic forms and have been the subject of intense research due to their potential to change numerous industries, including biomedical, aeronautical, and automotive. TPMS structures, such as the gyroid, Schwarz diamond, and Neovius surfaces, have various advantageous features. They have a high surface area-to-volume ratio, which improves thermal energy transmission and makes them suitable for applications such as heat exchangers and thermal insulators. TPMS structures can sustain significant mechanical loads while keeping a lightweight profile, which is very beneficial in aerospace and automotive applications where weight reduction is required without sacrificing strength. Certain TPMS structures are also biocompatible, making them appropriate for biomedical applications such as scaffolds for tissue engineering due to their porous architecture, which promotes cell development and nutrient flow.

AI can aid the design of Triply Periodic Minimal Surfaces (TPMS) by allowing for more efficient, precise, and novel ways. AI-driven design uses machine learning methods and computational methodologies to optimise the complicated geometries and features of TPMS structures. These algorithms can evaluate large quantities of data and learn from patterns, allowing for the development of TPMS designs that meet certain performance requirements such as mechanical strength, thermal conductivity, and biocompatibility. By automating the design process, AI saves time and effort when compared to traditional approaches, which frequently involve trial-and-error and expensive computational resources. AI may combine many design objectives and constraints at once, resulting in more resilient and multifunctional TPMS systems. Zhang et al. [8] investigated the elastic modulus of triply periodic minimal surface (TPMS) structures for titanium, a critical biomedical material, using three machine learning (ML) methods: Random Forest, XGBoost, and Adaboost. The researchers created a dataset from elastic finite element analysis of models with a large number of lattice-cells (4 × 4 × 4) and estimated elastic moduli based on unit configuration and two structural factors (k and C). The findings indicated good accuracy across all techniques, with Adaboost performing the best ($R^2$ = 0.959, MSE = 0.532) and Random Forest the lowest ($R^2$ = 0.929, MSE = 0.923). Hu et al. [9] studied into the optimization of triply periodic minimal surface (TPMS) designs for composite titania ceramics used in porous ceramic prosthetic bones. TPMS is praised for its superior bionic qualities and self-supporting capabilities. However, improving the TPMS design is difficult due to the extensive multi-parameter-property connections involved. In this study, a multi-objective optimization technique guided by finite element method (FEM) simulations was used to expedite the design process. Hussain et al. [10] focused on optimizing TPMS lattice architectures for 3D printing in order to increase production efficiency and reduce material waste. They provided a machine learning system for recommending optimal TPMS values based on engineering specifications. They examined four machine learning algorithms, K-Nearest Neighbors, Decision Tree, Random Forest, and Bayesian Regression, on a dataset of 144 polylactic acid (PLA) samples. The Random Forest and Decision Tree algorithms outperformed the others, with high R-squared values (0.9694 and 0.9689, respectively) and low RMSE (0.1180 and 0.0795). Ibrahimi et al. [11] tackled the problem of linking input parameters to mechanical and morphological features of Triply-Periodic Minimal Surfaces (TPMS) scaffolds. They generated a dataset of over 1,000 TPMS scaffolds and used Finite Element Modeling and image analysis to characterize them. Three machine learning (ML) models were trained, using both linear and non-linear approaches to predict input parameters. Feature selection for prediction was implemented in three ways: totally automatic (greedy algorithm), user-defined, and a hybrid of the two. Wang et al. [12] investigated the inverse design of shell-based mechanical metamaterials (SMM) modeled after Triply Periodic Minimal Surfaces (TPMS). Unlike previous research, which focused on predicting TPMS mechanical characteristics, this work sought to develop configurations based on unique loading curves, which is useful for applications such as energy absorption. The suggested approach combines machine learning (ML) for efficiency with genetic algorithms (GA) for global optimization. Wu et al. [13] developed an ML-based design technique to optimize ceramic additive manufacturing (AM) for functionally graded tissue scaffolds made of Triply Periodic Minimal Surfaces (TPMS). The goal was to meet the expected biomechanical criteria for bone regeneration. Their technique included a Bayesian optimization (BO) algorithm for time-dependent mechano-biological optimization of 3D printed ceramic scaffolds, resulting in high efficiency and cheap computing costs.

There is a need for an efficient and dependable approach that can successfully optimize the tensile stress behavior of TPMS architected materials, allowing for their broad use in a variety of engineering applications. This study proposes to overcome this difficulty by combining machine learning approaches with optimization algorithms. This study aims to create a strong computational framework capable of precisely predicting and optimizing the tensile stress performance of TPMS-architected materials, overcoming the limits of previous techniques.

## 2. Working Mechanism of Simulated Annealing

Simulated Annealing algorithm is based on random search method which was developed by Kirkpatrick et al. This algorithm is inspired by physical annealing process where materials are subjected to heat and then they are slowly cooled to remove defects and further achieve a state of minimum energy which allows downwards steps or transition to weaker solutions as well. So this algorithm is designed to find the global minimum of an objective function $f(x)$ by mimicking the physical annealing process. The main components of this algorithm are the objective function, the cooling schedule and the temperature. The probability of accepting the worse solutions is governed by the temperature $T$ which allows the solution to explore a boarder solution space by escaping from the local minima.

In this visualization shown in Figure 1, the landscape function is defined as a combination of sine and cosine functions with additional terms to create multiple peaks and valleys. The algorithm starts at a random point on the landscape and explores the space, gradually moving towards lower energy points. The progress of the algorithm is shown as red dots moving on the surface of the landscape.

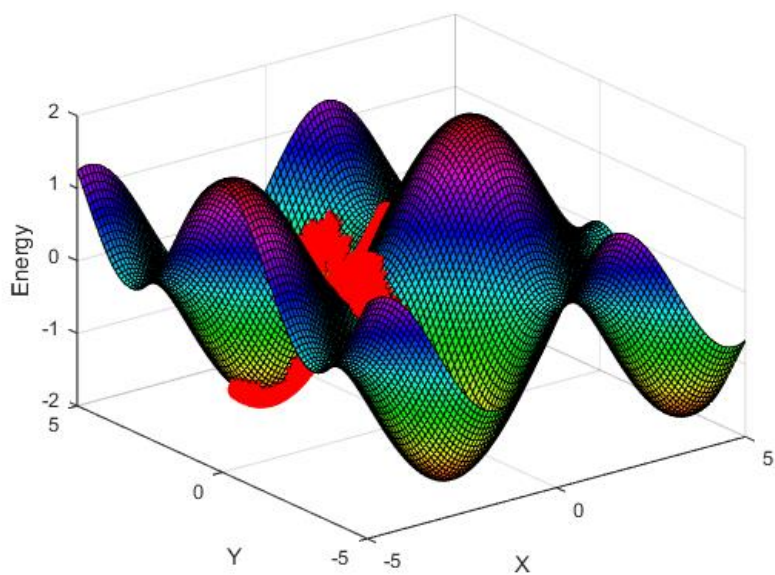

a)

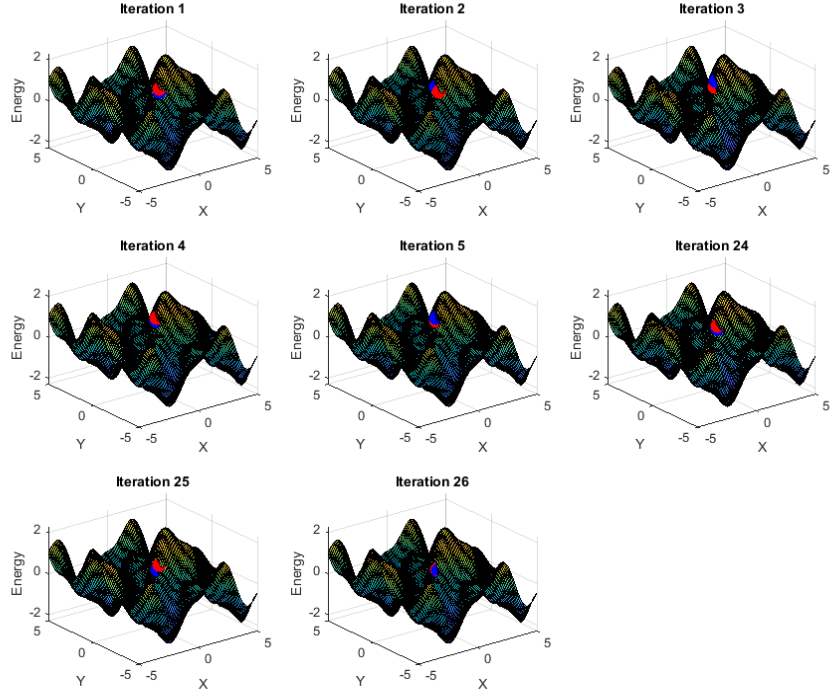

b)

**Figure 1.** a) Visualization of the Simulated Annealing (SA) algorithm applied to a landscape function. The landscape is defined by a combination of sine and cosine functions with multiple peaks and valleys, b) The algorithm begins at a random point and progresses towards lower energy points, depicted by red dots on the landscape surface.

The mechanism of the process initiates with an initial solution $x_0$ and an initial temperature $T_0$. By making a small random change to the current solution, a new candidate solution $x_{new}$ is generated as shown in equation 1.

$$x_{new} = x_0 + \Delta x \qquad (1)$$

The change in the objective function is calculated by using equation 2. If the new solution is better i.e. $\Delta E \leq 0$ then it is acceptable but if the new solution is worse i.e. $\Delta E \geq 0$ then it is accepted with a probability $P$ given by the Metropolis criterion shown in equation 3.

$$\Delta E = f(x_{new}) - f(x_{current}) \qquad (2)$$

$$P(accept) = e^{\left(-\frac{\Delta E}{T}\right)} \qquad (3)$$

It is clearly observed from the equation 3 that $P$ decreases as the $T$ decreases which reduces the acceptance of the worst solutions as the algorithm progresses. The temperature $T$ is gradually reduced according to a cooling schedule which can be exponential cooling as shown in equation 4.

$$T_{new} = \alpha T_{current} \qquad (4)$$

Where $\alpha$ is a constant less than 1. Repeating the process continues until a stopping requirement is satisfied, which could be lowering the temperature, finishing a predetermined number of iterations, or meeting an objective function threshold.

3. **Material and Methods**

In this study, five types of lattice structures were investigated: Gyroid, Fischer-Koch S, IWP, Schwarz D, and Karcher K, each designed with a relative density of 50%. The mechanical performance of these structures was evaluated using three metallic alloys: AA2024 T3, AISI304, and Ti6Al4V, selected for their high strength, corrosion resistance, and biocompatibility. Compression tests were conducted at pressures of 20 MPa, 30 MPa, 40 MPa, and 50 MPa, resulting in 60 simulations. Each lattice structure was modeled in a 4x4x4 configuration with a representative volume element (RVE) of 19.6 mm³, ensuring precise geometric parameters shiwn in Figure 2. Finite element analysis (FEA) was employed to simulate the mechanical behavior under these conditions, ensuring that each lattice structure was modeled accurately with the specified relative density for calculating the maximum tensile stress generated in lattice structures shown in Figure 3. Table 1 shows the input features and the output feature i.e. Tensile stress (MPa).

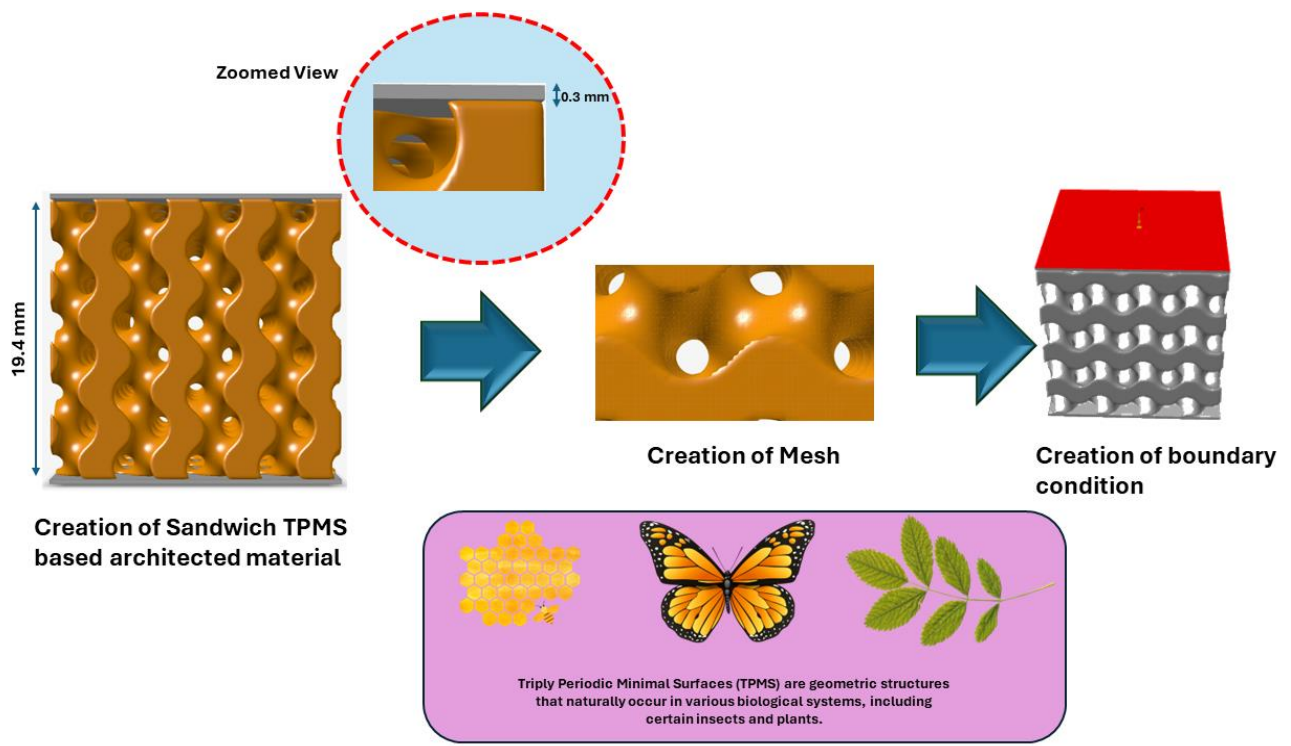

**Figure 2.** Steps involved in preparing the dataset using FEA

Table 1. Mechanical properties of different TPMS lattice types under varying applied pressures

| Lattice Type | Young Modulus of Alloy (GPa) | Poisson Ratio | Applied Pressure (MPa) | Tensile Stress (MPa) |
|---|---|---|---|---|
| Gyroid | 200 | 0.29 | 20 | 270.9 |
| Gyroid | 200 | 0.29 | 30 | 406.3 |
| Gyroid | 200 | 0.29 | 40 | 541.7 |
| Gyroid | 200 | 0.29 | 50 | 677.1 |
| Gyroid | 73.1 | 0.33 | 20 | 330.4 |
| Gyroid | 73.1 | 0.33 | 30 | 495.7 |
| Gyroid | 73.1 | 0.33 | 40 | 660.9 |
| Gyroid | 73.1 | 0.33 | 50 | 826.1 |
| Gyroid | 113.8 | 0.342 | 20 | 310.9 |
| Gyroid | 113.8 | 0.342 | 30 | 466.4 |
| Gyroid | 113.8 | 0.342 | 40 | 621.9 |
| Gyroid | 113.8 | 0.342 | 50 | 777.4 |
| Schwarz D | 200 | 0.29 | 20 | 131.9 |
| Schwarz D | 200 | 0.29 | 30 | 197.9 |
| Schwarz D | 200 | 0.29 | 40 | 263.8 |
| Schwarz D | 200 | 0.29 | 50 | 329.8 |
| Schwarz D | 73.1 | 0.33 | 20 | 228.9 |
| Schwarz D | 73.1 | 0.33 | 30 | 343.4 |
| Schwarz D | 73.1 | 0.33 | 40 | 457.8 |
| Schwarz D | 73.1 | 0.33 | 50 | 572.3 |
| Schwarz D | 113.8 | 0.342 | 20 | 185 |
| Schwarz D | 113.8 | 0.342 | 30 | 277.5 |
| Schwarz D | 113.8 | 0.342 | 40 | 370 |
| Schwarz D | 113.8 | 0.342 | 50 | 462.5 |
| IWP | 200 | 0.29 | 20 | 156.9 |
| IWP | 200 | 0.29 | 30 | 235.3 |
| IWP | 200 | 0.29 | 40 | 313.7 |
| IWP | 200 | 0.29 | 50 | 392.1 |
| IWP | 73.1 | 0.33 | 20 | 276 |
| IWP | 73.1 | 0.33 | 30 | 414 |
| IWP | 73.1 | 0.33 | 40 | 552 |
| IWP | 73.1 | 0.33 | 50 | 690 |
| IWP | 113.8 | 0.342 | 20 | 194.6 |
| IWP | 113.8 | 0.342 | 30 | 292 |
| IWP | 113.8 | 0.342 | 40 | 389.3 |
| IWP | 113.8 | 0.342 | 50 | 486.6 |
| Fischer Koch S | 200 | 0.29 | 20 | 100.3 |
| Fischer Koch S | 200 | 0.29 | 30 | 150.4 |
| Fischer Koch S | 200 | 0.29 | 40 | 200.5 |
| Fischer Koch S | 200 | 0.29 | 50 | 250.7 |
| Fischer Koch S | 73.1 | 0.33 | 20 | 102.6 |
| Fischer Koch S | 73.1 | 0.33 | 30 | 153.9 |
| Fischer Koch S | 73.1 | 0.33 | 40 | 205.2 |
| Fischer Koch S | 73.1 | 0.33 | 50 | 256.5 |
| Fischer Koch S | 113.8 | 0.342 | 20 | 107.9 |
| Fischer Koch S | 113.8 | 0.342 | 30 | 161.9 |
| Fischer Koch S | 113.8 | 0.342 | 40 | 215.9 |
| Fischer Koch S | 113.8 | 0.342 | 50 | 269.9 |
| Karcher K | 200 | 0.29 | 20 | 276.6 |
| Karcher K | 200 | 0.29 | 30 | 414.9 |
| Karcher K | 200 | 0.29 | 40 | 553.2 |
| Karcher K | 200 | 0.29 | 50 | 691.5 |
| Karcher K | 73.1 | 0.33 | 20 | 392.9 |
| Karcher K | 73.1 | 0.33 | 30 | 589.4 |
| Karcher K | 73.1 | 0.33 | 40 | 785.8 |
| Karcher K | 73.1 | 0.33 | 50 | 982.3 |
| Karcher K | 113.8 | 0.342 | 20 | 328.8 |
| Karcher K | 113.8 | 0.342 | 30 | 493.3 |
| Karcher K | 113.8 | 0.342 | 40 | 657.7 |
| Karcher K | 113.8 | 0.342 | 50 | 822.1 |

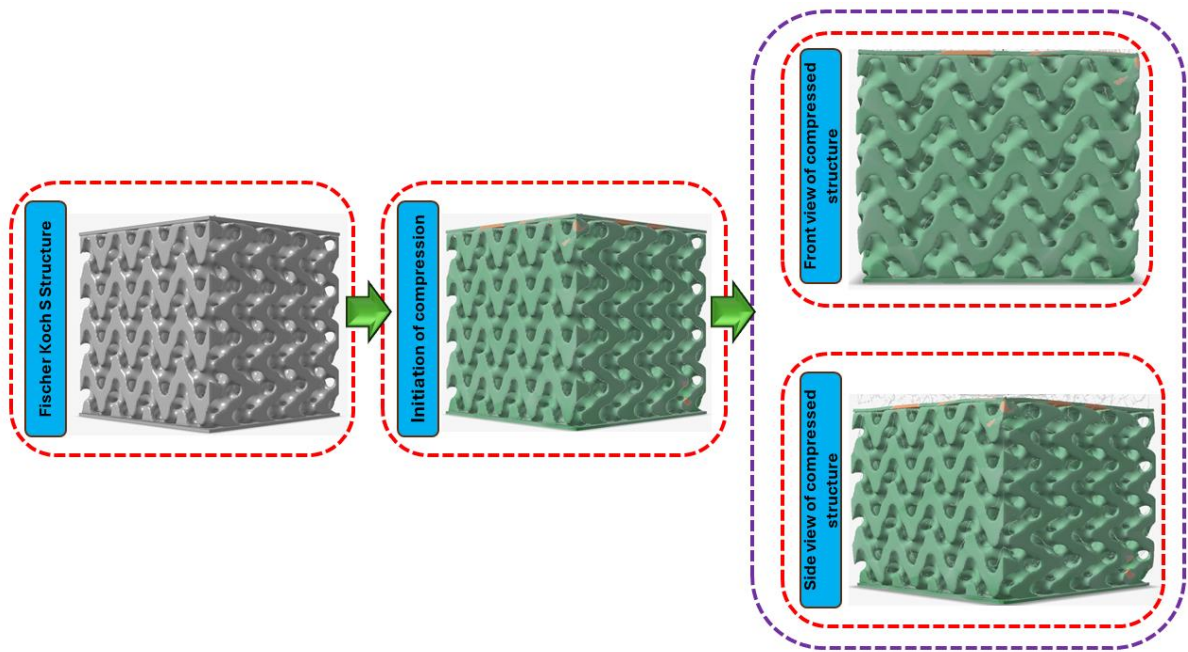

a)

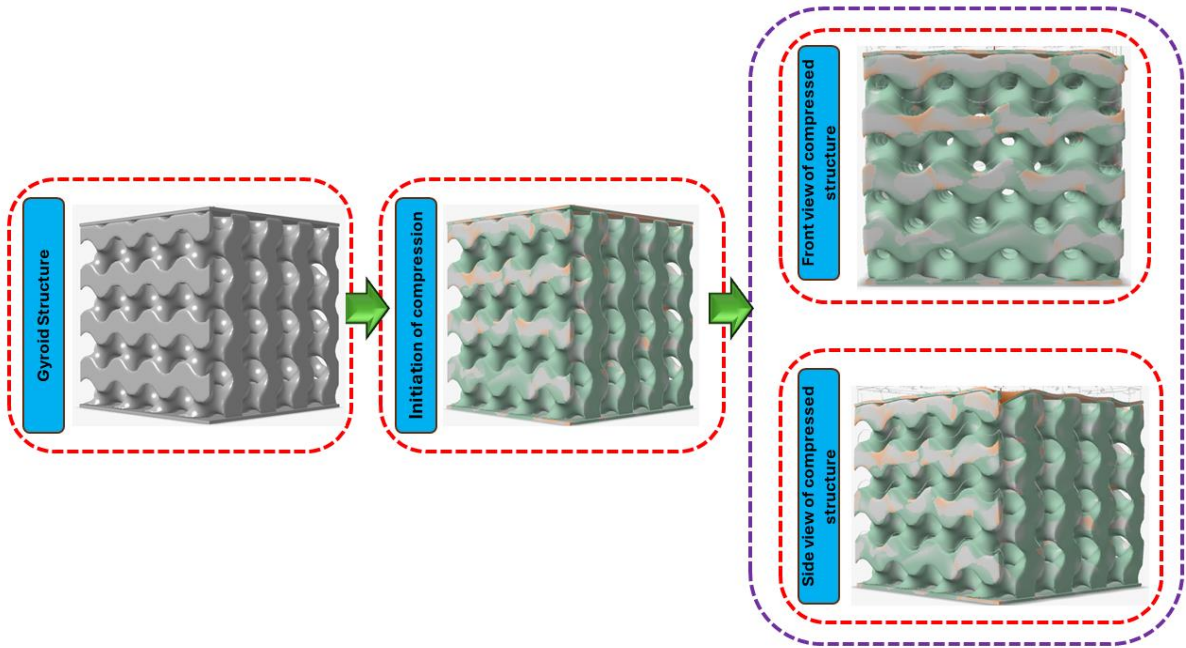

b)

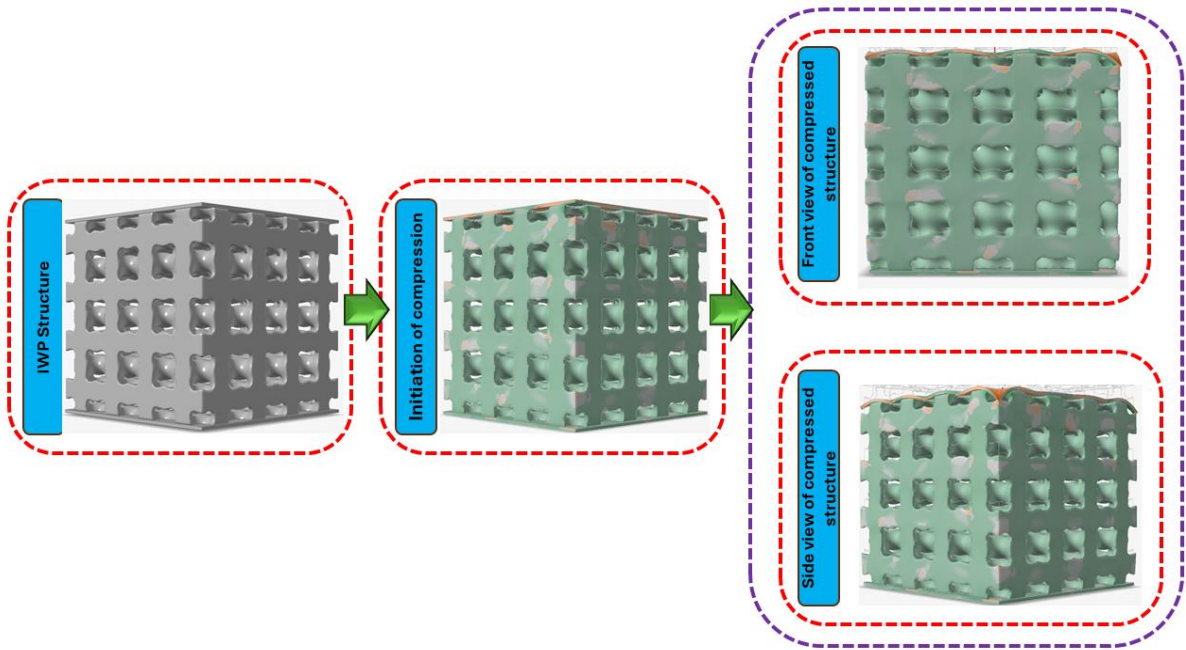

c)

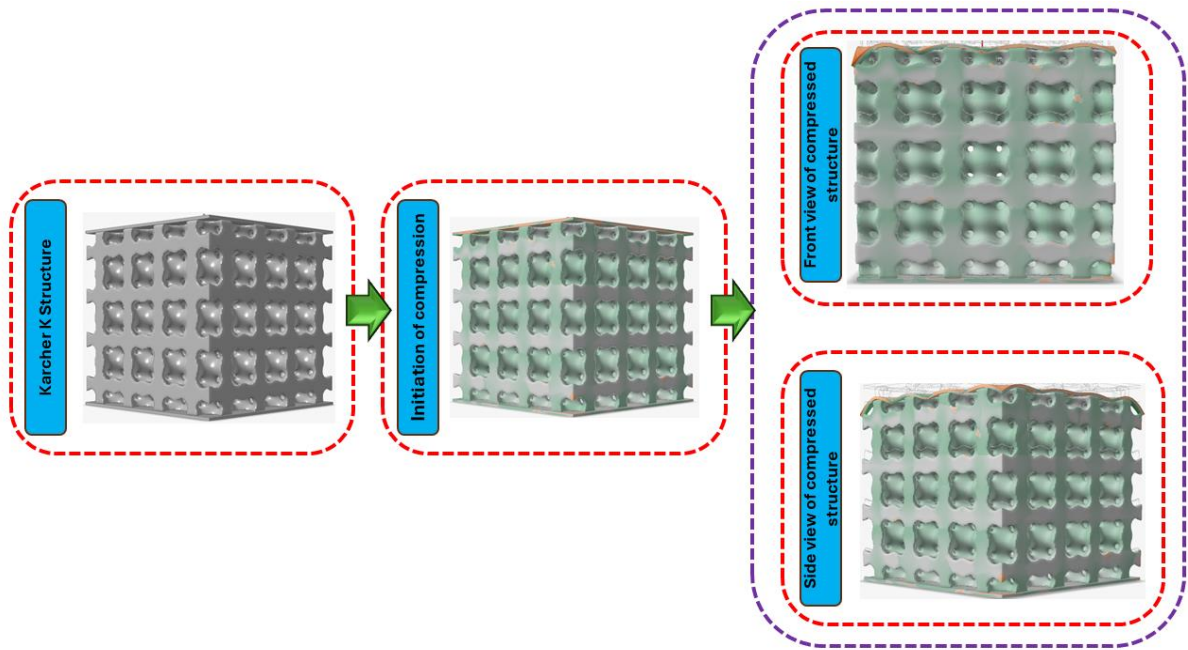

d)

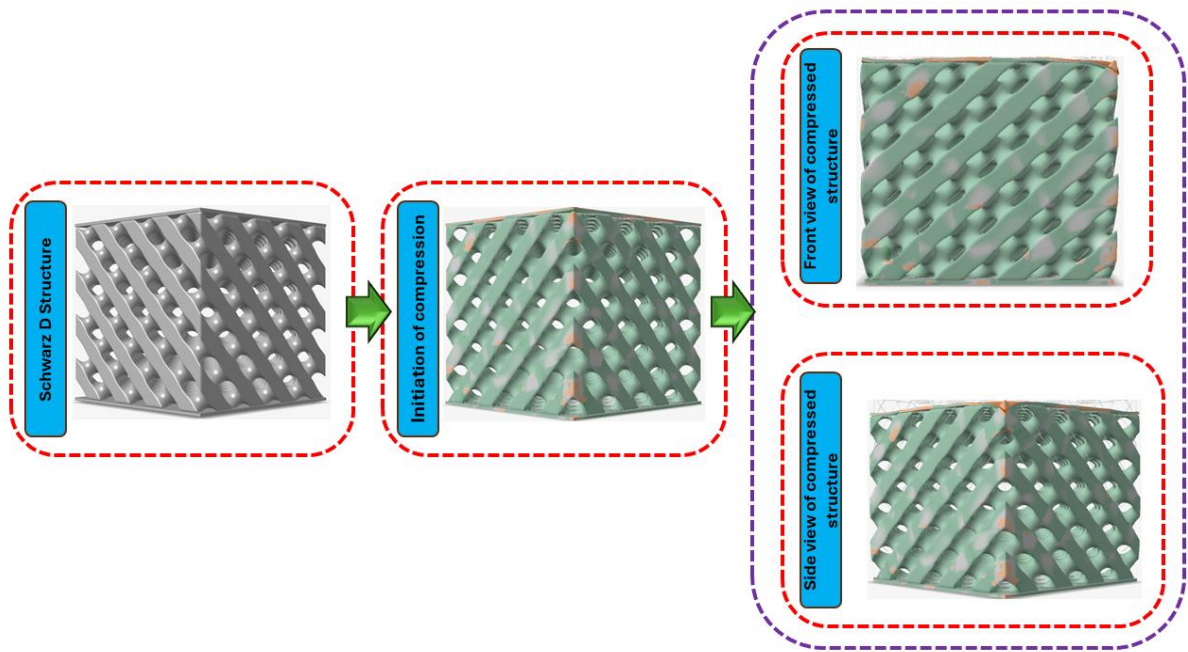

e)

**Figure 3.** Compression testing of a)Fischer Koch S, b) Gyroid, c) IWP, d) Karcher K and e) Schwarz D based architected materials at the compression pressure of 50 MPa

For the machine learning component, the dataset was split into features (X) i.e. Lattice type, Young's modulus of Alloy (GPa), Poison ratio of the alloy and the applied pressure (MPa) and the target variable (y), which was the tensile stress in MPa. The objective was to optimize the machine learning model's hyperparameters using Simulated Annealing. Initially, features and target variables were separated, and the data was split into training and validation sets in 80-20 ratio. The objective function aimed to minimize the negative R-squared value for the model's predictions on the validation set. Hyperparameters shown in Table 2 and Table 3 were optimized using Simulated Annealing, starting with an initial parameter set and iteratively cooling down the search temperature. The minimize function from the scipy.optimize library was employed with the Nelder-Mead method for optimization. The optimal model, with the best hyperparameters, was then evaluated for its performance using metrics such as Root Mean Squared Error (RMSE), Root Mean Absolute Error (RMAE), and R-squared ($R^2$).

Table 2. Hyperparameters used in SA-XGBoost algorithm

| Hyperparameter | Description | Initial Value |
|---|---|---|
| n_estimators | Number of boosting rounds | 100 |
| max_depth | Maximum tree depth for base learners | 10 |
| learning_rate | Boosting learning rate | 0.1 |
| min_child_weight | Minimum sum of instance weight (hessian) needed in a child | 1 |

Table 3. Hyperparameters used in SA-Random Forest and SA-Decision Tree algorithm

| Hyperparameter | Description | Initial Value |
|---|---|---|
| n_estimators | The number of trees in the forest | 100 |
| max_depth | The maximum depth of the tree | 10 |
| min_samples_split | The minimum number of samples required to split an internal node | 2 |
| min_samples_leaf | The minimum number of samples required to be at a leaf node | 1 |
| max_features | The number of features to consider when looking for the best split | auto |
| criterion | The function to measure the quality of a split (*gini* or *entropy*) | gini |

## 4. Results and Discussion

The tensile strength of a material is important in determining the performance and integrity of a lattice structure under compressive force. When compressive forces are applied to a lattice structure, different parts of the structure experience a variety of stresses, including compression and tension. This phenomena results from the lattice's unique geometry and load distribution channels. During compression, the lattice structure must be able to sustain both compressive and tensile loads caused by bending and deformation. The material's tensile strength guarantees that tensioned portions do not fail early. If the material's tensile strength is poor, these tensile zones may crack or fracture, even if the overall loading condition is compressive. The material's tensile strength helps to maintain the overall stability of the lattice structure during compression. By resisting tensile stresses that arise in specific locations, the material aids in the prevention of buckling, a major failure mode for lattice structures under compressive loads. The tensioned sections effectively offset the tendency of adjacent regions to bow, ensuring the lattice's structural integrity and stability.

The level of tensile stress experienced by regions within a lattice structure under compression is a critical factor in determining the structural integrity and potential for failure. High tensile stress implies that certain areas of the lattice are being subjected to significant pulling forces. If these tensile stresses exceed the material's tensile strength limit, it can lead to cracking or catastrophic failure, as materials have a finite capacity to withstand tensile loading before breaking. Consequently, high tensile stress in a lattice structure is generally undesirable, as it indicates that the structure is operating close to or beyond the material's ability to resist those stresses, posing a risk of structural failure. In contrast, a low tensile stress indicates that the forces pulling the material apart are well below acceptable limits. When tensile stresses are low, the likelihood of tensile failure, such as cracking or breaking, decreases. Under these conditions, the lattice structure can safely support the applied loads without exceeding the material's tensile strength. Low tensile stress is thus a desirable situation because it indicates that the structure is well within the material's acceptable operating limits, lowering the risk of failure due to excessive tensile loading. Table 4-7 shows the visualization of the stress generated in the different type of metallic based lattice structures under different compressive pressure.

**Table 4.** Visualization of generated stresses at compressive pressure of 20 MPa

| Lattice Type | AA2024 T3 | AISI 304 | Ti6Al4V |
|---|---|---|---|
| Fischer Koch S | 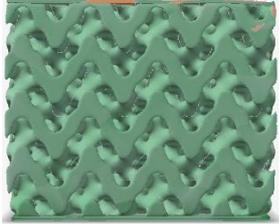 | 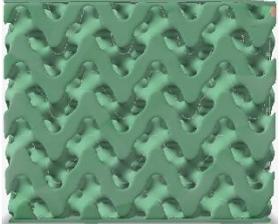 | 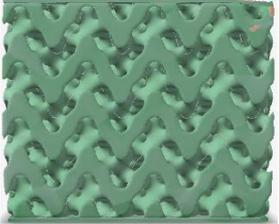 |
| Gyroid | 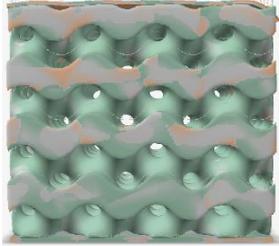 | 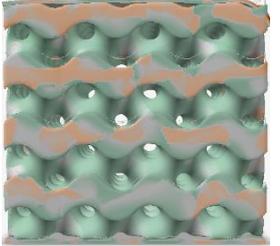 | 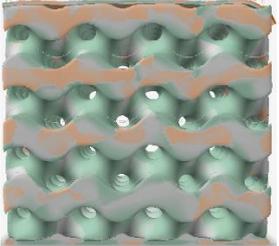 |
| Karcher K | 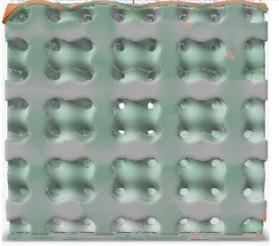 | 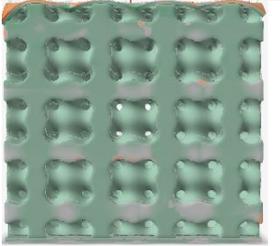 | 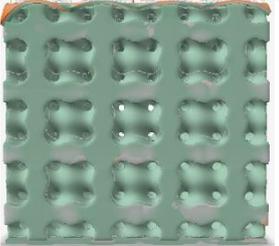 |

| Lattice Type | AA2024 T3 | AISI 304 | Ti6Al4V |
|---|---|---|---|
| IWP | 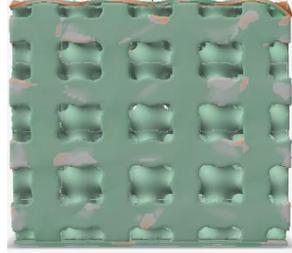 | 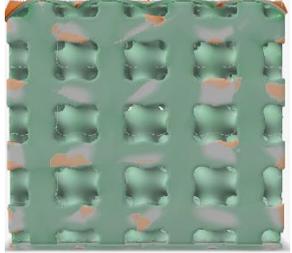 | 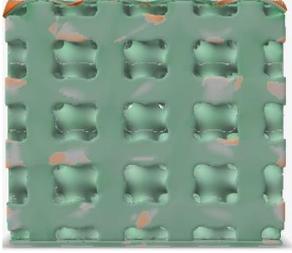 |
| Schwarz D | 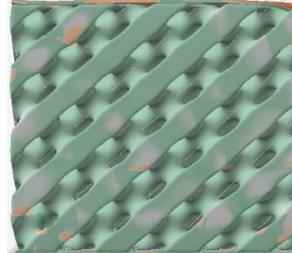 | 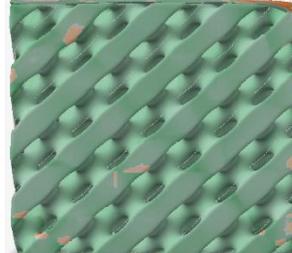 | 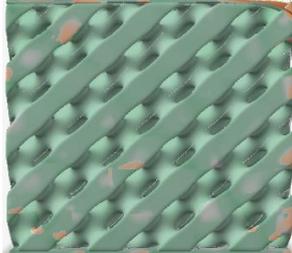 |

**Table 5.** Visualization of generated stresses at compressive pressure of 30 MPa

| Lattice Type | AA2024 T3 | AISI 304 | Ti6Al4V |
|---|---|---|---|
| Fischer Koch S | 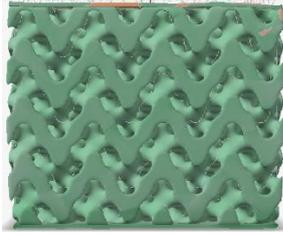 | 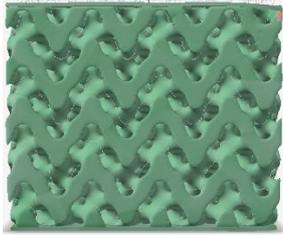 | 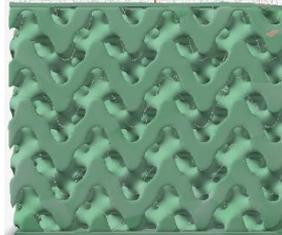 |
| Gyroid | 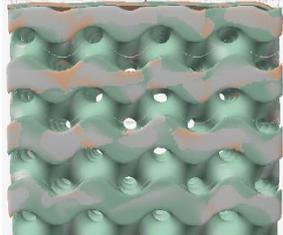 | 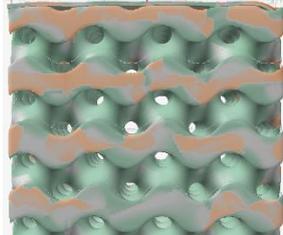 | 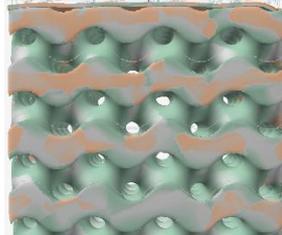 |
| Karcher K | 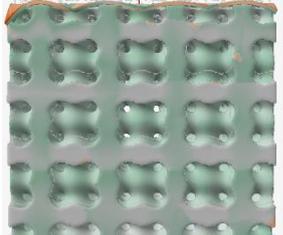 | 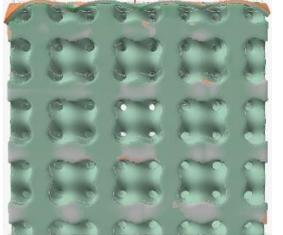 | 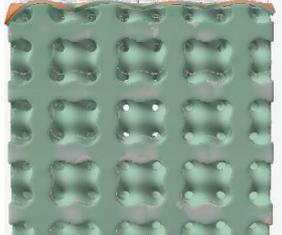 |

| Lattice Type | AA2024 T3 | AISI 304 | Ti6Al4V |
|---|---|---|---|
| IWP | 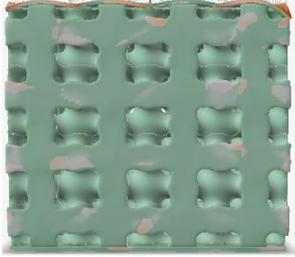 | 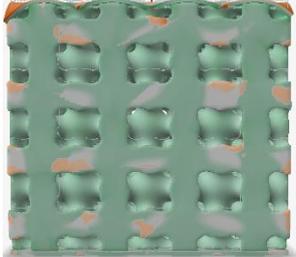 | 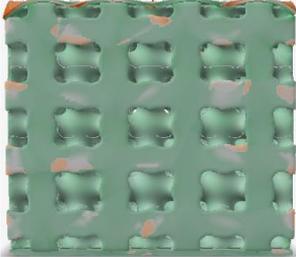 |
| Schwarz D | 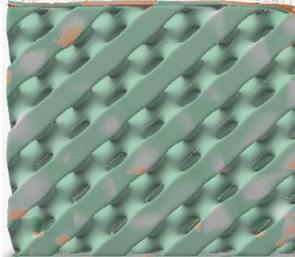 | 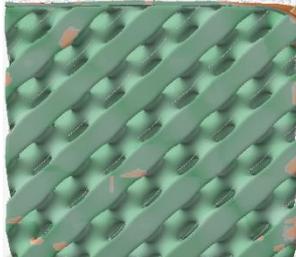 | 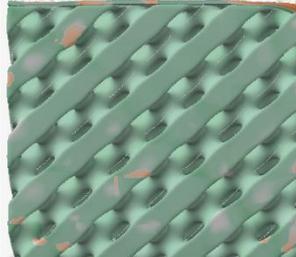 |

**Table 6.** Visualization of generated stresses at compressive pressure of 40 MPa

| Lattice Type | AA2024 T3 | AISI 304 | Ti6Al4V |
|---|---|---|---|
| Fischer Koch S | 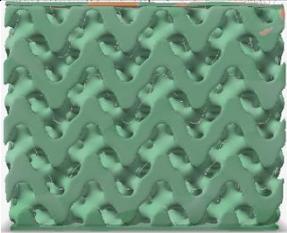 | 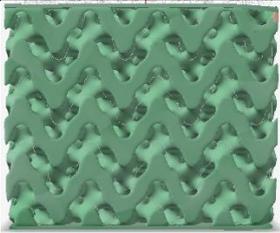 | 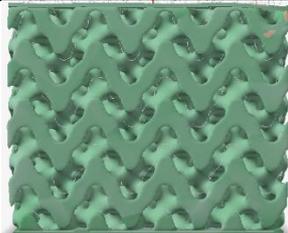 |
| Gyroid | 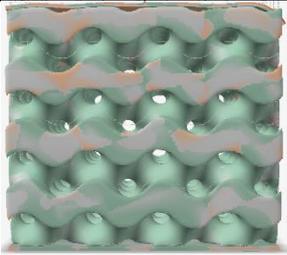 | 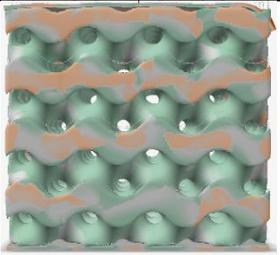 | 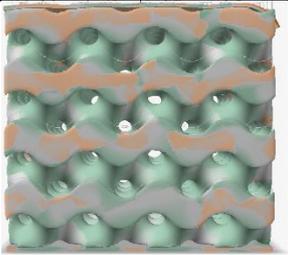 |
| Karcher K | 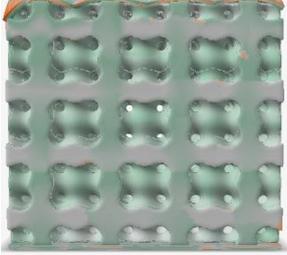 | 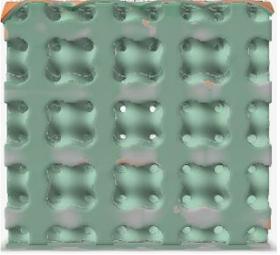 | 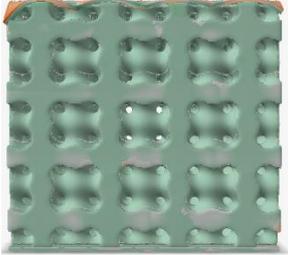 |

| Lattice Type | AA2024 T3 | AISI 304 | Ti6Al4V |
|---|---|---|---|
| IWP | 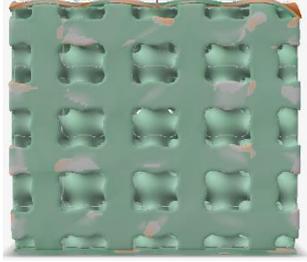 | 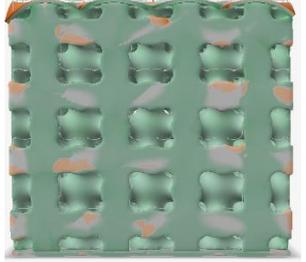 | 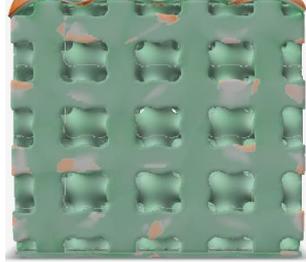 |
| Schwarz D | 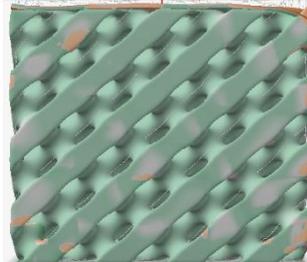 | 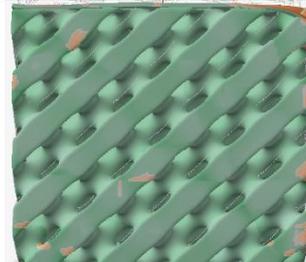 | 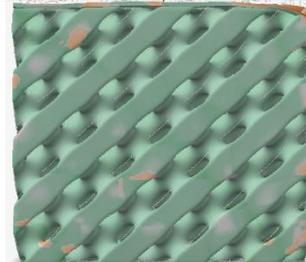 |

**Table 7.** Visualization of generated stresses at compressive pressure of 50 MPa

| Lattice Type | AA2024 T3 | AISI 304 | Ti6Al4V |
|---|---|---|---|
| Fischer Koch S | 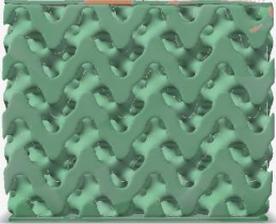 | 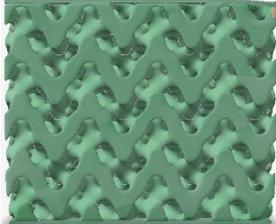 | 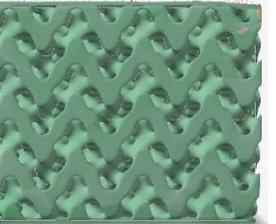 |
| Gyroid | 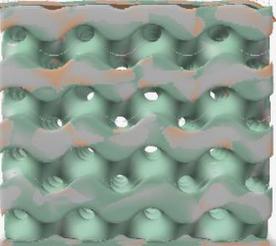 | 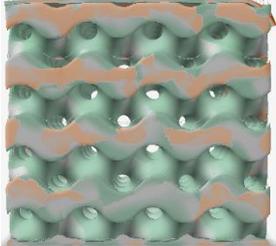 | 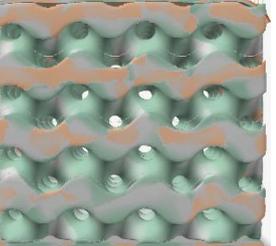 |
| Karcher K | 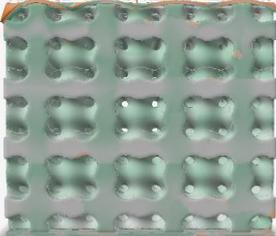 | 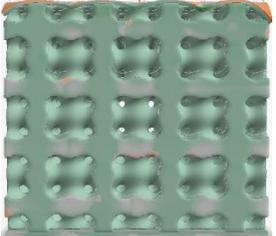 | 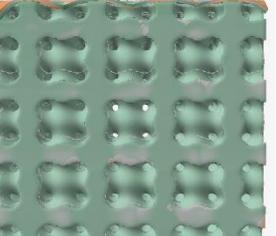 |
| IWP | 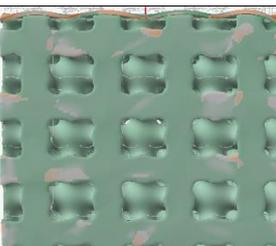 | 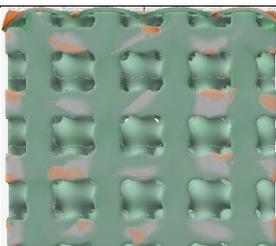 | 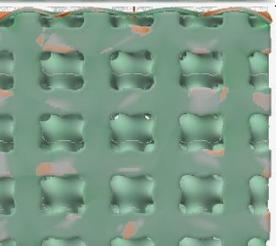 |
| Schwarz D | 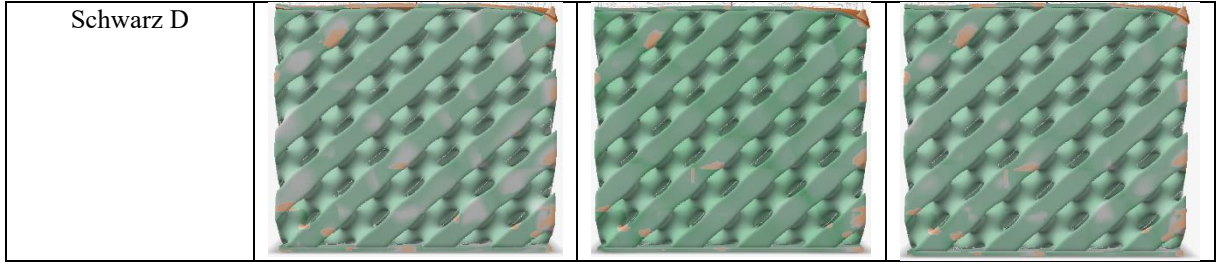 | | |

Integrating the supervised learning capability of ML algorithms with the stochastic optimization capabilities of SA allows for the coupling of SA with ML based regressors for hyperparameter optimization. The primary objective is to fine-tune the hyperparameters of the ML models to achieve optimal performance on a given dataset, specifically targeting parameters such as n_estimators, max_depth, learning_rate, and min_child_weight. This process begins with defining an objective function $f(p)$ that measures the performance of the model. In this context, the negative R-squared score on the validation set is used as the objective function to be minimized shown in equation 5.

$$f(p) = -R^2.p \qquad (5)$$

Where $p$ = {$n_{estimators}$, max_depth, learning_rate, min_child_weight}.

The Simulated Annealing algorithm is initialized with a starting set of hyperparameters and an initial temperature, $T_0$. During the optimization process, new candidate solutions are generated by making small random perturbations to the current hyperparameter values. These new candidates are then evaluated using the objective function, resulting in a performance metric. The acceptance of the new solution is decided by the difference in performance $\Delta E$ which is given by the equation 6.

$$\Delta E = f(p_{new}) - f(p_{current}) \qquad (6)$$

The new solution is accepted if the condition $\Delta E \leq 0$ is satisfied as it improves the objective function. If $\Delta E > 0$ then the new solution is accepted with a probability $P(accept)$. The optimization process iteratively adjusts the hyperparameters, balancing exploration and exploitation through the probabilistic acceptance mechanism of SA. This allows the algorithm to escape local minima and explore a broader solution space, increasing the likelihood of finding the global optimum. The process continues until a stopping criterion is met, such as reaching a predefined number of iterations or achieving a sufficiently low temperature. Once the SA algorithm has identified the appropriate hyperparameters, the training set is used to train the specific ML model with these parameters. The performance of the optimized model is then assessed on the validation set using measures such as Mean Squared Error (MSE), Mean Absolute Error (MAE), and R-squared score. These metrics give a complete picture of the model's prediction accuracy and reliability. The combination of SA and a specific ML model thus takes advantage of the benefits of both techniques, yielding a powerful approach to hyperparameter optimization that improves the model's performance on regression tasks.

The learning curves in the graphs depicted in Figure 4 show the performance and generalization ability of three different machine learning models: SA-Random Forest, SA-Decision Tree, and SA-XG Boost. Each plot depicts the training (red line) and validation (green line) scores as a function of the number of training samples. The major purpose of these plots is to assess and compare the models' ability to learn from and generalize to new data.

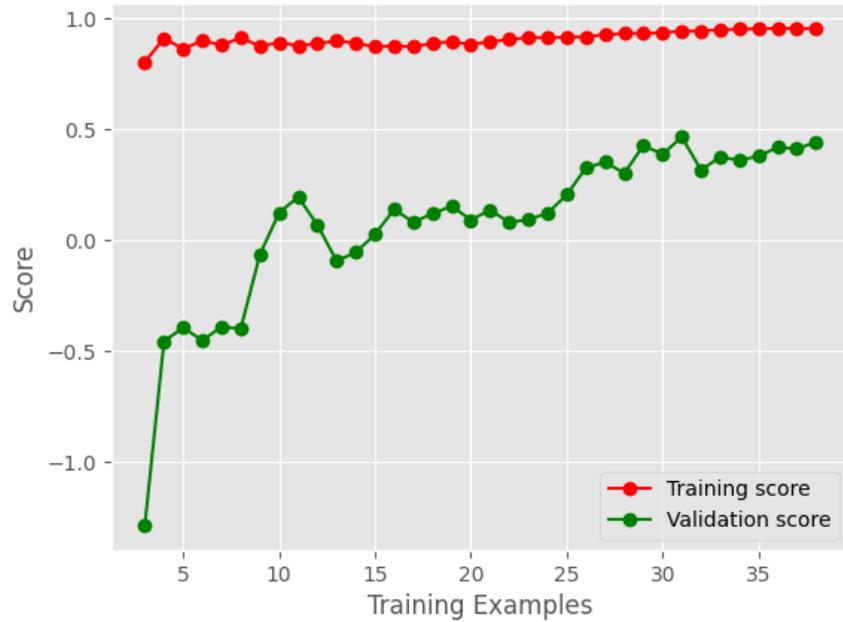

a)

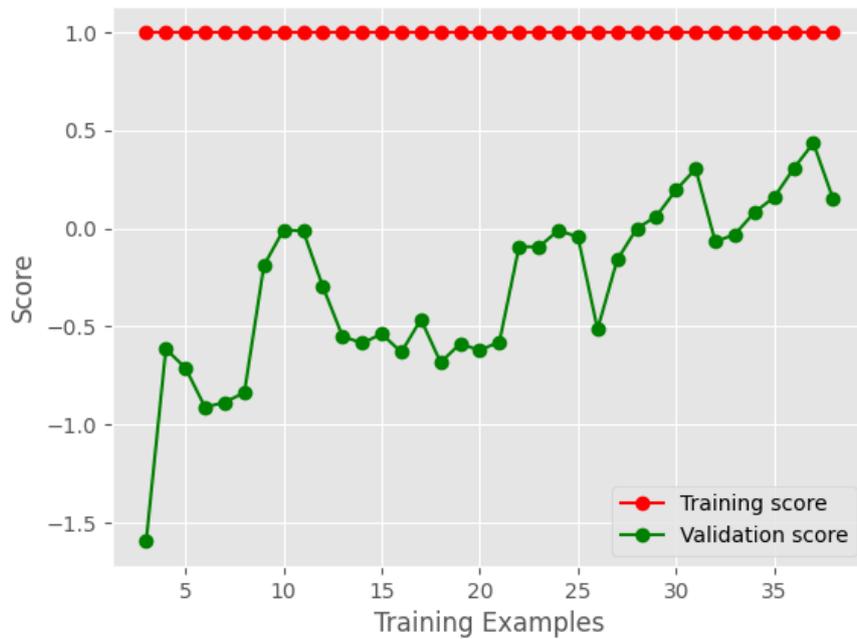

b)

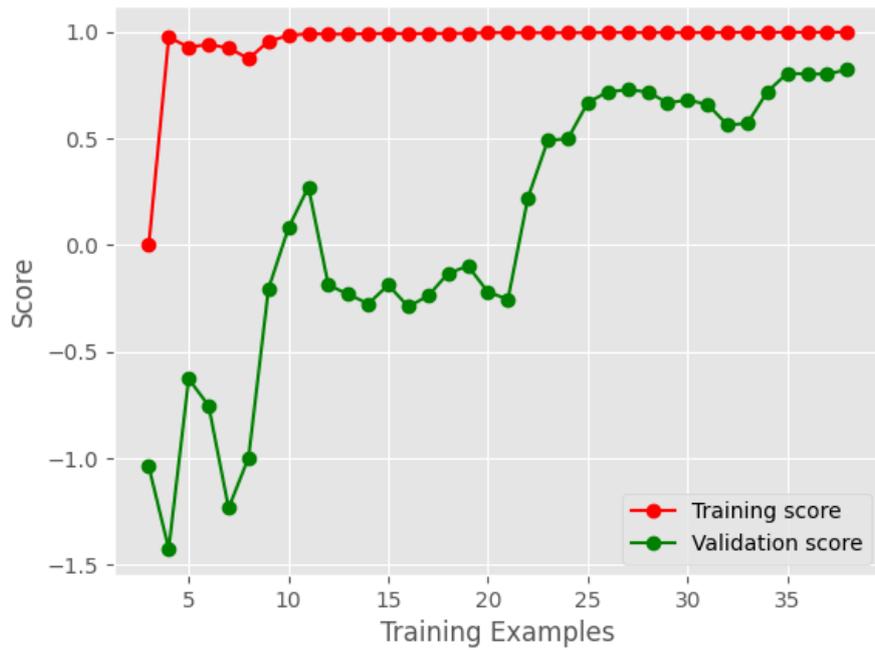

c)

**Figure 4.** Learning curves for a) SA-Random Forest, b) SA-Decision Tree and c) SA-XGBoost

The SA-Random Forest model's training score is constantly high, near to one, suggesting that the model fits the training data extraordinarily well. This high training score indicates that the Random Forest model is successfully capturing patterns in the training data. However, the validation score begins low and progressively improves as the number of training examples rises, eventually stabilizing at 0.5. The increasing validation score suggests that when more data is used, the model's ability to generalize to previously unseen data improves. Nonetheless, the difference between the training and validation scores suggests some overfitting, which is common with Random Forest models due to their inherent complexity.

The SA-Decision Tree model has a similar high training score, indicating a strong match with the training data. However, the validation score for this model is initially low and fluctuates more than the Random Forest, eventually settling around 0.2. The high training score indicates that the Decision Tree fits the training data very well, however the lower and more variable validation score reveals that the Decision Tree is more prone to overfitting, particularly with smaller quantities of data. The changes in the validation score suggest that the Decision Tree is extremely sensitive to data perturbations, which results in inconsistent performance on the validation set.

The SA-XGBoost model also has a continuously high training score, near to one, indicating excellent fit to the training data. The validation score for XGBoost begins low but gradually climbs, eventually settling around 0.8. The high training score indicates that XGBoost is very good at detecting patterns in the training data. The continually increasing validation score implies that XGBoost is more generalizable than the Decision Tree and Random Forest models.

A smaller difference between training and validation scores indicates less overfitting and improved generalization ability.

The validation curves depicted in Figure 5 show the performance of three different machine learning models—SA-Random Forest, SA-Decision Tree, and SA-XGBoost—at various maximum depth parameter values. Each plot shows the training score (red line) and validation score (green line) as a function of maximum depth, which determines the model's complexity.

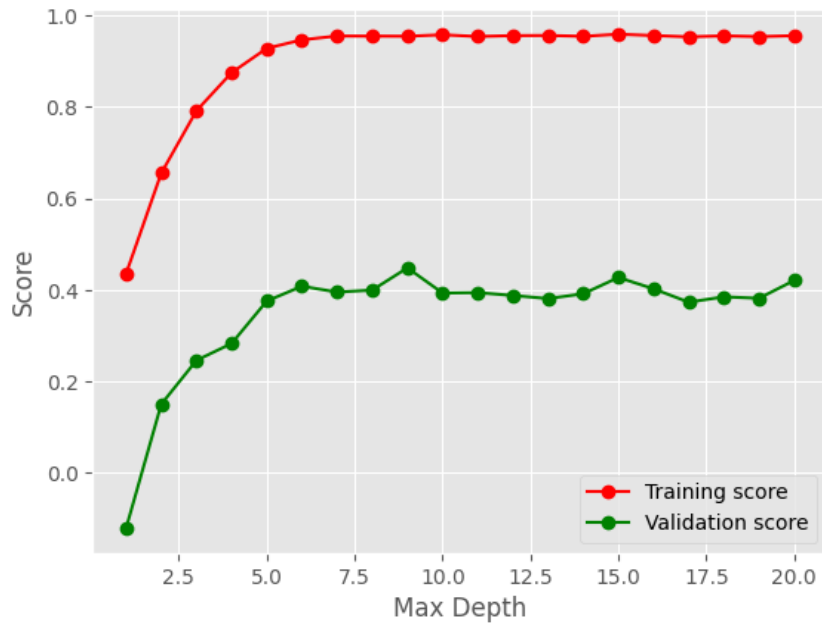

a)

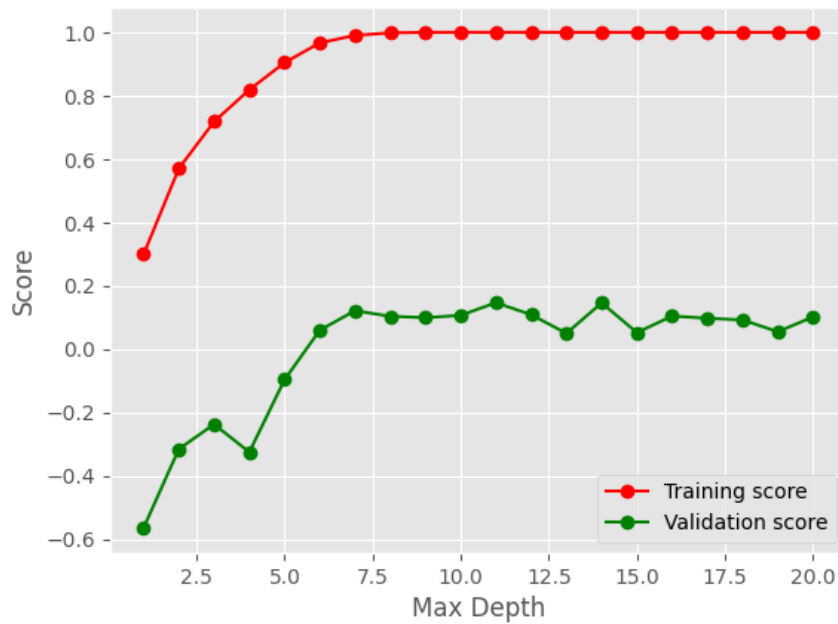

b)

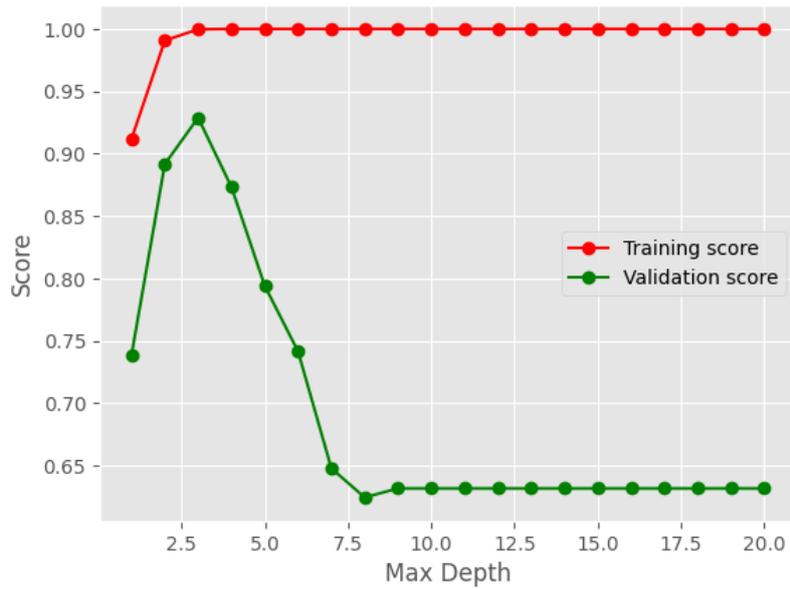

c)

**Figure 5.** Validation curves for a) SA-Random Forest, b) SA-Decision Tree and c) SA-XGBoost

The training score for the SA-Random Forest model is continuously high, reaching 1 at all maximum depth settings. This high training score suggests that the model accurately captures patterns in the training data, regardless of depth. The validation score begins low and rises as the maximum depth increases, eventually settling around 0.5 at higher depths. This trend indicates that the Random Forest model requires enough depth to reflect the complexity of the data, and once it does, it performs better on the validation set. The diminishing difference between training and validation scores at greater depths suggests less overfitting and better generalization ability.

The SA-Decision Tree model follows a similar trend, with the training score remaining high at all depths, indicating a strong match to the training data. However, the validation score for this model fluctuates significantly, especially at lower levels, before stabilizing around 0.1 at higher depths. The initial strong increase in the validation score, followed by oscillations, indicates that the Decision Tree model is more sensitive to changes in depth. The model requires a certain depth to perform properly on unseen data, but after that point, further depth has no significant effect on performance. The persistent high training score, along with changing validation values, shows a larger proclivity for overfitting as compared to the Random Forest.

The SA-XGBoost model, like the other models, has a high training score across all depths, indicating good fit to the training data. The validation score begins high, peaking around a depth of 3, and then gradually decreases as the maximum depth increases, eventually stabilizing around 0.65. This initial surge followed by a fall shows that XGBoost can better capture data complexity at lower depths than the other models. However, as the depth grows, the model may get overfit, resulting in a poorer validation score. The initial high validation score and subsequent drop illustrate the importance of carefully setting the maximum depth parameter to strike a balance between model complexity and generalization capacity.

The plots shown in Figure 6 visualize the difference between the predicted values and the actual values (residuals) for each data point in the dataset.

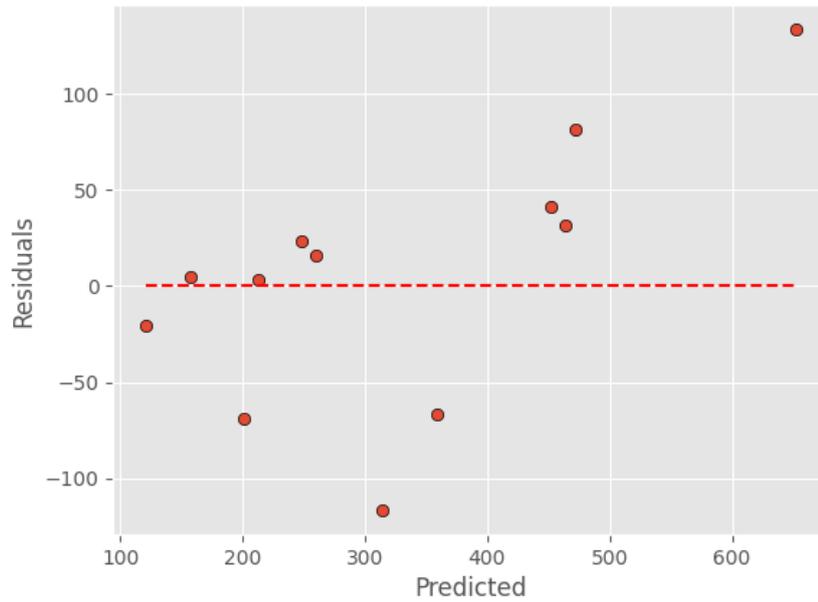

a)

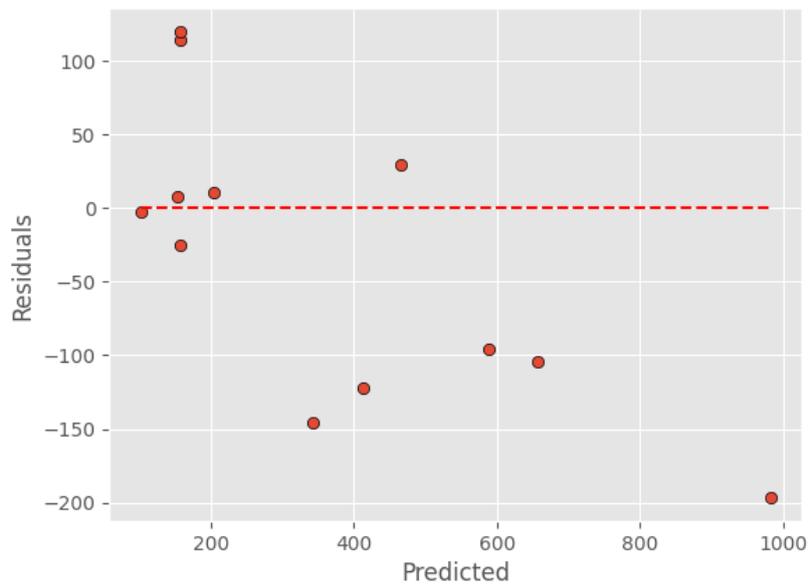

b)

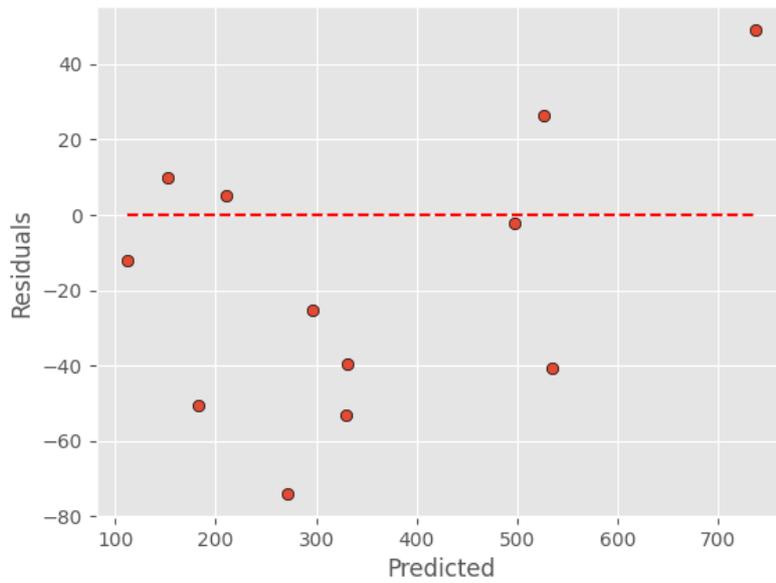

c)

**Figure 6.** Residual plots for a) SA-Random Forest, b) SA-Decision Tree and c) SA-XGBoost

In the Random Forest residuals plot, the residuals are rather uniformly spread around the zero line, indicating that the model fits the data well. While there are some outliers, the majority of the residuals are clustered within a respectable range, indicating that the Random Forest model does a fair job of forecasting the target variable. The Decision Tree residuals plot displays a more scattered pattern with greater residuals, especially in the higher predicted value range. This shows that the Decision Tree model may struggle to effectively anticipate the target variable, particularly for larger values. The presence of bigger residuals suggests a poorer fit than the Random Forest model. Finally, the XGBoost residuals illustration shows a tighter distribution of residuals near the zero line than the other two models. The residuals are often less in magnitude, indicating that the XGBoost model fits the data more accurately. This shows that the XGBoost model is likely the best-performing of the three, as it can predict the target variable more accurately.

The results of the Simulated Annealing (SA) optimized machine learning models are summarized in the provided Figure 7 and Table 8, depicting the performance of three different algorithms: Random Forest, Decision Tree, and XGBoost. The performance metrics considered include Root Mean Squared Error (RMSE), Root Mean Absolute Error (RMAE), and the coefficient of determination ($R^2$). The SA-Random Forest model demonstrated a robust performance with an RMSE of 65.38, an RMAE of 7.12, and an $R^2$ value of 0.89. These results indicate that the Random Forest model, when optimized using Simulated Annealing, can accurately predict the target variable, exhibiting a strong correlation between predicted and actual values as illustrated by the scatter plot. The data points are closely aligned along the diagonal line, suggesting minimal prediction error. In contrast, the SA-Decision Tree model, while still effective, showed a higher RMSE of 101.57 and an RMAE of 9.00, along with a lower $R^2$ value of 0.74. This suggests that the Decision Tree model had more significant discrepancies between predicted and actual values. The scatter plot further confirms this, showing a wider spread of data points around the diagonal line, indicating higher prediction errors compared to the Random Forest model. The SA-XGBoost model outperformed both the Random Forest and Decision Tree models, achieving the lowest RMSE of 38.81 and RMAE of 5.68, along with the highest R² value of 0.96. This superior performance is visually supported by the scatter plot, where the data points closely cluster around the diagonal line, reflecting high prediction accuracy and a very strong correlation between predicted and actual values.

**Table 8.** Depiction of the metric features for SA-ML models

| Algorithms | RMSE | RMAE | $R^2$ Value |
|---|---|---|---|
| SA-Random Forest | 65.38 | 7.12 | 0.89 |
| SA-Decision Tree | 101.57 | 9.00 | 0.74 |
| SA-XG Boost | 38.81 | 5.68 | 0.96 |

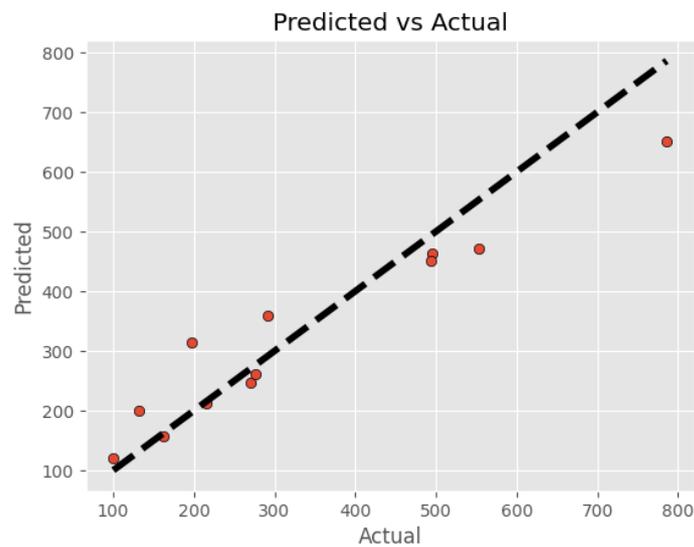

a)

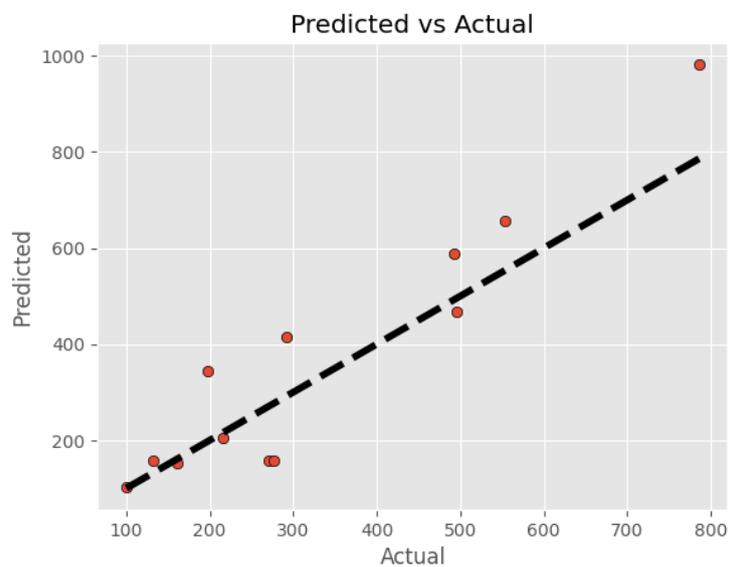

b)

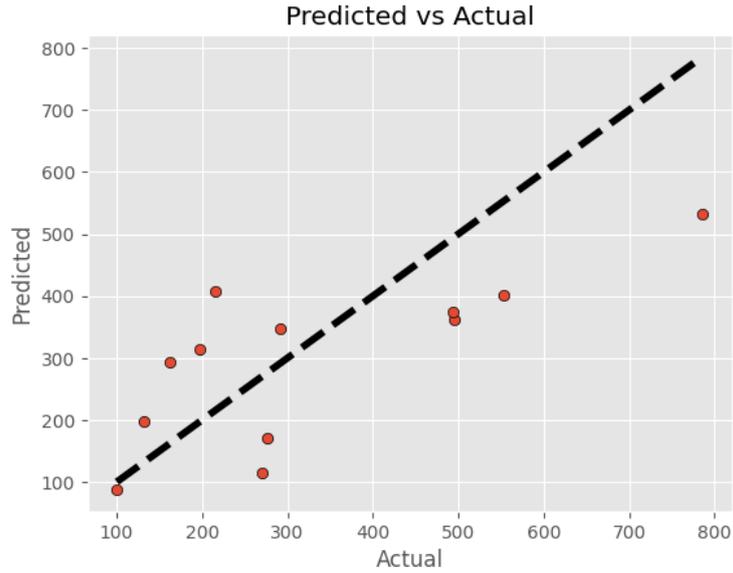

c)

**Figure 7.** Actual vs predicted values plots for a) SA-Random Forest, b) SA-Decision Tree and c) SA-XGBoost

## 5. Conclusion

This study presents a novel approach to optimizing the tensile stress of Triply Periodic Minimal Surface (TPMS) architected materials by integrating machine learning models with the Simulated Annealing (SA) optimization algorithm. The performance of three machine learning models, Random Forest, Decision Tree, and XGBoost, was evaluated for predicting tensile stress using a dataset generated from finite element analysis of TPMS models. The results demonstrated that the SA-XGBoost model outperformed the other models, achieving the highest coefficient of determination ($R^2$) value of 0.96, the lowest root mean squared error (RMSE) of 38.81, and the lowest root mean absolute error (RMAE) of 5.68. The SA-Random Forest model also exhibited promising performance with an $R^2$ value of 0.89, while the SA-Decision Tree model exhibited more fluctuations in validation scores and lower overall performance. The integration of Simulated Annealing with machine learning models proved effective in optimizing the hyperparameters, thereby enhancing the predictive capabilities of these models. The SA-XGBoost model demonstrated superiority in capturing the complex relationships within the data, making it the most suitable choice for predicting tensile stress in TPMS architected materials. While this study focused on predicting tensile stress, future research could explore the optimization of other mechanical properties of TPMS architected materials, such as compressive strength, fatigue life, and fracture toughness, using similar machine learning and optimization techniques.

**Conflict of interest**

Author declares there is no conflict of interest.

**Funding**

This research did not receive any external funding.

**Data availability statement**

The data that support the findings of this study are available upon request from the author.